\documentclass[11pt]{article}
\usepackage{amssymb,amsfonts,amsmath}
\usepackage[pdftex]{graphicx}
\DeclareGraphicsExtensions{.pdf}  \textheight24.5cm
\begin{document}

\renewcommand{\thefootnote}{\fnsymbol{footnote}}

\centerline{\Large \bf On the limits of the Navier--Stokes equations} 
\vspace*{3mm} \centerline{Peter Stubbe\footnote[1]{retired from Max--Planck--Institut f\"ur Sonnensystemforschung, 37077 G\"ottingen, Germany. Contact by e-mail: peter-stubbe@t-online.de }} 

\renewcommand{\thefootnote}{\arabic{footnote}}

\vspace*{.5cm}{\small \begin{quote} Heuristic derivations of the Navier--Stokes equations are unable to reveal the applicability limits of these equations. In this paper we rederive the Navier--Stokes equations from kinetic theory, using a method that affords a step by step insight into the required simplifying assumptions. The major task on this way is to find the conditions needed to truncate the resulting infinite system of transport equations at a finite level. The minimal obtainable closed set comprises three equations, for the particle number density $\mathit{N} $, the macroscopic velocity $\mathbf{v}$, and the temperature $\mathit{T}$. It is verified that this 3--equation system conserves the total energy, i.e., the sum of kinetic and internal energy. As a consequence, the energy is not conserved if the integrity of this closed system is violated, as for instance in the case of the so--called incompressible Navier--Stokes equations where the equation for $\mathbf{v}$ is the only one kept, and the other two discarded and jointly replaced by the incompressibility condition $\nabla \cdot \mathbf{v} = 0$. It is shown that a second viscosity, found in parts of the literature, does not exist as long as the Navier--Stokes equation is applied inside the range of its validity. Outside this range, a second viscosity builds up, however not as a matter constant. The Navier--Stokes equation in its known form rests upon the tacit assumption that the particles are points without volume and without collective forces between them, whereby dense gases and liquids are excluded, and the applicability limited to ideal gases. In the final section of this paper, an attempt is made to generalize the equations for applicability to real fluids. 
\end{quote} }

\vspace*{15mm} 
{\large \bf 1. Introduction}

The conversion of kinetic theory into transport theory gives rise to an infinite number of hierarchically ordered macroscopic transport equations. In order to obtain a tractable finite set, it is necessary to find truncation conditions allowing to disregard moments beyond a certain order. It turns out that the minimal achievable set comprises three equations, for the particle number density $N$, the macroscopic velocity {\bf v}, and the temperature $T$ (see Refs.~1--3). The appertaining truncation conditions set limits to the applicability of these transport equations. A well--known condition is the Knudsen condition, demanding that the mean free path is short compared with a typical scale of the process. However, this condition alone is by far not sufficient. 
 
We proceed as follows: In Sections 2 to 6 the equations belonging to the Navier--Stokes system will be rederived, using a method which affords a transparent view into the simplifications required to arrive at the known results. This will lead to a set of applicability conditions, allowing to decide whether a given mathematical solution is physically valid. A summarizing compilation of these conditions will be given in Section 8. In Section 7 it will be verified that the Navier--Stokes system conserves the total energy, i.e., the sum of kinetic and internal energy, and expressions will be given for the various energy transfer rates involved in the total energy balance. A critical discussion of the physical consequences of the frequently used incompressibility assumption will follow in Section 9. The Navier--Stokes equation in its known form rests upon the assumption the the particles are points without own volume and without intermolecular forces between them. In Section 10, an attempt will be made to incorporate the effects of the finite size of particles and the collective forces between them, in order to extend the validity of the Navier--Stokes equations beyond ideal gases to dense gases and liquids.

\newpage

\vspace*{5mm} {\large \bf 2. Basic equations}

The mother equation for the system of macroscopic transport equations is the kinetic equation, describing the distribution function in phase space, $f({\bf u}, {\bf r}, t)$. The meaning of $f$ is that $f\, d^3{\bf u}\, d^3{\bf r}$ gives the number of particles within the phase space volume element $d^3{\bf u}\,d^3{\bf r}$, so that $ \int \! f \, d^3{\bf u} = N$ is the particle number density. 

Considering a one--constituent neutral fluid without particle production or loss, the kinetic equation is given by

\begin{equation} 
\frac{d\, f}{d\, t} + ({\bf u} - {\bf v})\cdot \nabla f + \frac{{\bf F}}{m} \cdot \nabla_u f = \frac{\delta f}{\delta t}
\end{equation}

where $df/dt = \partial f/dt \, + \, {\bf v} \cdot \nabla f$, {\bf v} is the macroscopic velovity, $\delta f/\delta t$ symbolically denotes the temporal change of $f$ due to the action of collisions, {\bf F} is a force acting on the particles individually, and $\nabla_u $ is the nabla operator in velocity space. The force {\bf F} may depend on {\bf r} and $t$, but not on {\bf u} (unless {\bf F} is perpendicular to {\bf u}).  Furthermore it will be assumed that collisions are sufficiently soft to be non--ionizing, $\delta \! \int \! f d^3 {\bf u} / \delta t = \delta N/\delta t = 0$. 

The common way to derive macroscopic equations from the kinetic equation (1), with $\delta f / \delta t$ expressed by the Boltzmann collision integral, is to appoximate the distribution function $f$ by an expansion about the Maxwellian 

\begin{equation} 
f = \frac{N}{(2 \pi)^{3/2}\,V^3}\,\, {\rm exp} \bigg[\! - \frac{({\bf u}-{\bf v} )^2}{2V^2}\, \bigg]
\end{equation}

where $V$ is the thermal velocity ($V^2 = KT/m$), with $K$ Boltzmann's constant and $m$ the particle mass. The two leading methods in this field are those of Chapman and Enskog (see Ref.~1) and Grad$^2$. A comprehensive review is given by Schunk$^3$.

The general problem in transforming the kinetic equation into macroscopic transport equations, independent of the method used, is that any moment of $f$ of given order $n$ is coupled with moments of order $n+1$, so that an infinite system of transport equations arises. The major task, therefore, will be to find conditions allowing to truncate the infinite sytem at a finite level in order to obtain a usable closed set of equations. 

The truncation conditions set limits to the applicability of the resulting equations. To establish these applicability conditions, we will not follow the methods of Chapman--Enskog or Grad. For the given purpose, it will be more appropriate to stay entirely within the frame of transport equations, without knowledge of the distribution function. The price to be paid is that the occurring transport coefficients -- viscosity and heat conductivity -- will have to be determined by use of a collisional relaxation model. This price, however, does not appear too high in view of the fact that in practical applications of the transport equations empirical values   for these transport coefficients will be used. 

We define a complete set of moments of $f$ by

\begin{equation} 
M^{(i,j,k)}({\bf r}, t) =  \int f({\bf u}, {\bf r}, t) \, \phi^{(i,j,k)}({\bf u}) \, d^3{\bf u}
\end{equation}
with 
\begin{equation} 
\phi^{(i,j,k)}({\bf u}) = m \, (u_x - v_x)^i \, (u_y - v_y)^j \, (u_z - v_z)^k
\end{equation}
and
\begin{equation} 
{\bf v} = \frac{1}{N}\int f\, {\bf u}\, d^3{\bf u}
\end{equation}

where $i$, $j$, $k$ are non--negative integers, and $n = i+j+k$ is the order of the moment. Multiplication of the terms in (1) by $\phi^{(i,j,k)}$ with subsequent intergration over velocity space yields, after some elementary mathematical steps,

\begin{equation} 
{\rm T1} + {\rm T2} + {\rm T3} + {\rm T4} = 0
\end{equation}

with terms T1 to T4 given by

\begin{equation} 
{\rm T1} = \frac{dM^{(i,j,k)}}{dt} + M^{(i,j,k)}\Big( \,\nabla\cdot {\bf v} + i\,\frac{\partial v_x}{\partial x} + j\,\frac{\partial v_y}{\partial y} + k\,\frac{\partial v_z}{\partial z} \Big)  - \frac{\delta M^{(i,j,k)}}{\delta t}
\end{equation}

\begin{equation} 
{\rm T2} =  i\, M^{(i-1,j,k)}\,\Big(\frac{dv_x}{dt}-\frac{F_x}{m} \Big) +  j\, M^{(i,j-1,k)}\,\Big(\frac{dv_y}{dt}-\frac{F_y}{m} \Big) + k\, M^{(i,j,k-1)}\,\Big(\frac{dv_z}{dt}-\frac{F_z}{m} \Big) 
\end{equation}

\vspace*{-3mm}
\begin{eqnarray} 
{\rm T3} =  \mbox{\hspace{-7mm}} & &  i\, \Big(M^{(i-1,j+1,k)}\,\frac{\partial v_x}{\partial y} + \,M^{(i-1,j,k+1)}\,\frac{\partial v_x}{\partial z} \Big) +
 j\, \Big(M^{(i+1,j-1,k)}\,\frac{\partial v_y}{\partial x} + \,M^{(i,j-1,k+1)}\,\frac{\partial v_y}{\partial z} \Big) + \nonumber \\
& &  k\, \Big(M^{(i,j+1,k-1)}\,\frac{\partial v_z}{\partial y} +  \, M^{(i+1,j,k-1)}\,\frac{\partial v_z}{\partial x} \Big)   
\end{eqnarray} 

\vspace*{-2mm}
\begin{equation} 
{\rm T4} = \frac{\partial M^{(i+1,j,k)}}{\partial x} + \frac{\partial M^{(i,j+1,k)}}{\partial y} + \frac{\partial M^{(i,j,k+1)}}{\partial 	  z}
\end{equation}

For the conversion of the last left--hand side term in (1) into a moment term it has been necessasry to postulate that the decrease of $f$ is stronger than the increase of $\phi^{(i,j,k)}$ as $|\bf{u}| $ goes to infinity.

The term T1 contains the wanted moment $ M^{(i,j,k)}$ of order $n = i+j+k$, the term T2 moments of order $n-1$, the term T3 other moments of order $n$, and the term T4 moments of order $n+1$. It is this fourth term that reveals the dilemma of transport theory. 

Before making use of (6)--(10), we identify the moments of orders $n=0$ to $3$.

For $n=0$: \vspace*{4mm}
\begin{equation} 
M^{(0,0,0)} = Nm \mbox{\hspace*{8mm} ({\rm mass density}) }
\end{equation}

For $n=1$: \vspace*{4mm}
\begin{equation} 
M^{(1,0,0)} = M^{(0,1,0)} = M^{(0,0,1)} = 0
\end{equation}

For $n=2$: \vspace*{4mm}
\begin{equation} 
M^{(2,0,0)} = p_{xx} \mbox{\hspace{4mm},\hspace{4mm}} M^{(0,2,0)} = p_{yy}
\mbox{\hspace{4mm},\hspace{4mm}} M^{(0,0,2)} = p_{zz}
\end{equation}
\vspace*{-2mm}
\begin{equation} 
M^{(1,1,0)} = p_{xy} = p_{yx}\mbox{\hspace{4mm},\hspace{4mm}} M^{(1,0,1)} =
p_{xz} = p_{zx} \mbox{\hspace{4mm},\hspace{4mm}} M^{(0,1,1)} = p_{yz} = p_{zy}
\end{equation}
\vspace*{-2mm}
\begin{equation} 
p = \frac{1}{3}\, (M^{(2,0,0)} + M^{(0,2,0)} + M^{(0,0,2)}) \mbox{\hspace*{8mm} ({\rm hydrostatic pressure}) }
\end{equation}
\begin{equation} 
T=\frac{1}{3KN}\, (M^{(2,0,0)} + M^{(0,2,0)} + M^{(0,0,2)}) \mbox{\hspace*{8mm} ({\rm kinetic temperature}) }
\end{equation}

For $n=3$: \vspace*{4mm}
\begin{equation} 
q_x = \frac{1}{2}\,( M^{(3,0,0)} + M^{(1,2,0)} + M^{(1,0,2)})
\end{equation}
\vspace*{-2mm}
\begin{equation} 
q_y = \frac{1}{2}\,( M^{(2,1,0)} + M^{(0,3,0)} + M^{(0,1,2)})
\end{equation}
\vspace*{-2mm}
\begin{equation} 
q_z = \frac{1}{2}\,( M^{(2,0,1)} + M^{(0,2,1)} + M^{(0,0,3)})
\end{equation}

These are the three components of the heat flux vector {\bf q}. 

The moment equations (6)--(10) in their complete form are unusable in practice because there are infinitely many. Let us speak of fluid theory when the infinite set of moment equations is reduced to a finite set. According to Stubbe$^4$, a reduction of the infinite set of moment equations to a finite set of fluid equations is possible if the condition

\begin{equation} 
\frac{1}{\tau^2} + \frac{1}{\tau_c^2} \gg \frac{V^2}{l^2}
\end{equation}

is fulfilled, where $\tau$ is a characteristic time and $l$ a characteristic length of the process under consideration, and $\tau_c$ the average time between collisions. Condition (20) comprises two limiting cases, viz.~the inertial fluid limit

\begin{equation} 
\tau \ll \tau_c \mbox{\hspace*{5mm} {\rm and} \hspace*{5mm}} \tau V \ll l
\end{equation}

and the collisional fluid limit 

\begin{equation} 
\tau_c \ll \tau \mbox{\hspace*{5mm} {\rm and} \hspace*{5mm}} \tau_c V \ll l
\end{equation}

Condition (21) typically applies to plasmas, whereas (22) is better suited for application to neutral fluids which are usually collision dominated. The second part of (22) corresponds to the Knudsen condition, and the first part of (22) is the precondition for the Knudsen condition. It must be realized that condition (20) merely justifies the general applicability of fluid theory. Specific applications, like the Navier--Stokes equations, will require further applicability conditions.

\vspace*{5mm} {\large \bf 3. Transport equations for the fluid variables $N$, {\bf v} and $T$}

Eqs.~(6)--(10), together with the moment definitions (11) to (19), will now be used to extract equations for $N$, {\bf v} and $T$. We obtain:

For $n=0$: \vspace*{4mm}
\begin{equation} 
\frac{dN}{dt} = - N \, \nabla \cdot {\bf v}
\end{equation}

For $n=1$: \vspace*{4mm}
\begin{equation} 
\frac{d\bf v}{dt} = \frac{{\bf F}}{m} - \frac{1}{Nm} \left[\, \nabla \, p  +  \nabla \cdot (\textsf{p})^{\rm o}\, \right]
\end{equation}

where $(\textsf{p})^{\rm o}$ is the traceless part of the pressure tensor $\textsf{p}$, with non--diagonal elements defined by (14), and diagonal elements $p_{xx} - p$, $p_{yy} - p$ and $p_{zz} - p$ defined by (13) and (15). 
Combination of (15) and (16) yields

\begin{equation} 
p = NKT
\end{equation}

so that (24) can be written as 

\begin{equation} 
\frac{d\bf v}{dt} = \frac{{\bf F}}{m} - V^2 \left[\, \frac{\nabla T}{T} + \frac{\nabla N}{N} \, \right]  - \frac{1}{Nm} \nabla \cdot (\textsf{p})^{\rm o}
\end{equation}

For $n=2$: \vspace*{4mm}
\begin{equation} 
\frac{d p}{d t} = - \frac{5}{3} \, p \, \nabla \cdot {\bf v} - \frac{2}{3}\,\left[ \, (\textsf{p})^{\rm o} : \nabla {\bf v} +
\nabla \cdot {\bf q} \, \right]
\end{equation}

or, using (23) and (25),

\begin{equation} 
\frac{d T}{d t} = - \frac{2}{3} \, T \, \nabla \cdot {\bf v} - \frac{2}{3KN}\,\left[ \, (\textsf{p})^{\rm o} : \nabla {\bf v} +
\nabla \cdot {\bf q} \, \right]
\end{equation}

where $\nabla {\bf v}$ denotes the dyadic product. 

The derivation of these relations from (6)--(10) is straightforward and without approximations. Thus, with (6)--(10) given, the sytem [(23),(26),(28)] is exact, however useless at this stage since $(\textsf{p})^{\rm o}$ and {\bf q} are undetermined. 

The reason why we see no collision terms in the above transport equations for $N$, {\bf v} and $T$ is because we have assumed that collisions are non--ionizing ($ \delta N/\delta t = 0$), and because there exists no second constituent with which momentum and energy could be exchanged ($ \delta {\bf v}/\delta t = 0$ and $ \delta T/\delta t = 0$ or $ \delta p/\delta t = 0$). However, the randomizing action of collisions will play a dominant role in the treatment of $(\textsf{p})^{\rm o}$ and ${\bf q}$.

The transport equations (26) and (28) are open--ended since the moments $(\textsf{p})^{\rm o}$ and ${\bf q}$ establish a link to an infinite number of higher order moments. It will be necessary, therefore, to find criteria allowing to express $(\textsf{p})^{\rm o}$ and ${\bf q}$ in terms of $N$, {\bf v} and $T$, in order to obtain a closed system of transport equations with no other variables than these.  This task will be addressed in Section 5.

\vspace*{5mm} {\bf \large 4. The Euler system}

If the dominant role of randomizing collisions goes to the extreme that the system is completely isotropized, the $(\textsf{p})^{\rm o}$ and ${\bf q}$ terms disappear, and we have

\begin{equation} 
\frac{d \bf v}{d t} = \frac{{\bf F}}{m} - V^2 \left[\, \frac{\nabla T}{T} + \frac{\nabla N}{N} \, \right]  
\end{equation}

\begin{equation} 
\frac{d T}{d t} = - \frac{2}{3} \, T \, \nabla \cdot {\bf v}
\end{equation}

The system [(23),(29),(30)] may be termed the Euler system because the central equation (29) corresponds to the Euler equation.

The Euler system is complete, insofar as it comprises three equations for the three fluid variables $N$, {\bf v} and $T$. In large parts of the literature, however, the Euler system is used in a highly reduced form by omitting (23) and (30) and replacing both by $\nabla \cdot {\bf v}  = 0 $ (see the review by Constantin$^5$). It is hard to see how these so--called ``incompressible Euler equations'' should be able to generate physically significant results (cf.~Section 9).

Occasionally one finds the statement that the Euler equation is a special case of Newton's second law, with the acting force expressed  in terms of a pressure gradient, ${\bf F} = -(1/N)\nabla p$. One has to realize, however, that in a dynamic system with internal motions the pressure has to be taken as what it is, namely a 9--element tensor, with non--diagonal elements unequal zero and diagonal elements unequal among each other. It will require a thorough justification, therefore, to reduce the full pressure tensor $\textsf{p}$ to its isotropic limit $p\,${\sf U} (where {\sf U} is the unit tensor).

\vspace*{5mm} 
{\bf \large 5. Approximations for $(\textsf{p})^{\rm o}$ and {\bf q}}

In order to extend the Euler system to a corresponding Navier--Stokes system, we are left with the task to express $(\textsf{p})^{\rm o}$ and ${\bf q}$ in terms of $N$, {\bf v} and $T$. This will not be possible if all terms in (26) and (28) are treated as being on equal footing. It will be necessary to set up a hierarchy of terms, and this is done by treating as minor terms those which constitute the difference between (26) and (29), and between (28) and (30). 

With respect to the $(\textsf{p})^{\rm o}$ term, this means that $|\nabla \cdot (\textsf{p})^{\rm o}|$ has to be much smaller than $|\nabla p|$, resulting in the condition \vspace*{3mm}
\begin{equation} 
|p_{ij}| \ll p_{ii} \mbox{\hspace*{5mm} {\rm and} \hspace*{5mm}} |p_{ii} - p| \ll p
\end{equation}

For the {\bf q} term in (27) it follows that $|\nabla \cdot {\bf q}|$ must be much smaller than $|p\, \nabla \cdot {\bf v}|$, with the consequence that the term T4 (which contains the elements of the heat flux vector) has to be ignored when eqs.~(6)--(10) are applied for $n=2$. In other words, at this level of approximation $(\textsf{p})^{\rm o}$ will not depend on {\bf q}.
 
In order to determine $(\textsf{p})^{\rm o}$ for use in (26) and (28), we apply (6)--(10) to the pressure tensor elements $p_{xx}$ ($i=2, j=k=0$) and $p_{xy}$ ($i=j=1, k=0$). The term T2 vanishes due to (12). Since $\delta p/\delta t = 0$, the collision term $\delta p_{xx}/\delta t$ in T1 can be altered into $\delta(p_{xx} - p)/\delta t$, and from (6) with ${\rm T2} = {\rm T4} = 0$ we obtain

\begin{equation} 
\frac{\delta(p_{xx} - p)}{\delta t}\, = \, \frac{dp_{xx}}{dt} + p_{xx} \left( \nabla\cdot {\bf v} + 2\, \frac{\partial v_x}{\partial x} \right) + 2\, p_{xy} \frac{\partial v_x}{\partial y} +2\, p_{xz} \frac{\partial v_x}{\partial z}
\end{equation}

Applying (31) to (32), we can ignore the last two terms and replace on the right--hand side $p_{xx}$ by $p$.  Then, expressing $dp/dt$ by the leading term in (27) ($dp/dt = - (5/3)\, p \, \nabla \cdot {\bf v} $), and writing $p=NKT$ according to (25), we obtain

\begin{equation} 
\frac{\delta(p_{xx} - p)}{\delta t}\, =\, NKT \left( 2\, \frac{\partial v_x}{\partial x} - \frac{2}{3}\, \nabla\cdot {\bf v} \right)
\end{equation}

With arguments corresponding to those leading to (33), $\delta p_{xy}/\delta t$ is obtained as

\begin{equation} 
\frac{\delta p_{xy}}{\delta t}\, = \, NKT \left( \frac{\partial v_x}{\partial y} + \frac{\partial v_y}{\partial x} \right)
\end{equation}

Expressions corresponding to (33) and (34) are obtained for the other elements of the traceless pressure tensor, and the results can be written in compact form as

\begin{equation} 
\frac{\delta(\textsf{p})^{\rm o}}{\delta t} =  NKT\, \left[\, \nabla {\bf v} + (\nabla {\bf v})^t - \frac{2}{3}\, (\nabla\cdot {\bf v}) \, \textsf{U} \, \right]
\end{equation}

where $(\nabla {\bf v})^t$ denotes the transposed dyad.

To finalize the derivation, we need a relation between $\delta(\textsf{p})^{\rm o}/\delta t $ and $(\textsf{p})^{\rm o}$. For this, we apply the 10--moment relaxation model of Stubbe$^6$, which is a considerable extension of the frequently used model by Gross and Krook$^7$, and we obtain

\begin{equation} 
\frac{\delta(\textsf{p})^{\rm o}}{\delta t} =  - \, 2\, \nu^{(r)} (\textsf{p})^{\rm o}
\end{equation}

where the collision frequency $\nu^{(r)}$ is a measure of the strength of the randomizing action of collisions, quantitatively defined by eqs.~(28c) and (15) of Ref.~6. The other two collision frequencies involved in the model, $\nu^{(m)}$ and $\nu^{(e)}$, describing momentum and energy transfer, do not appear here due to the absence of a second constituent.

From (35) and (36)

\begin{equation} 
(\textsf{p})^{\rm o} =  - \,  \eta \left[\, \nabla {\bf v} + (\nabla {\bf v})^t - \frac{2}{3}\, (\nabla\cdot {\bf v}) \, \textsf{U} \, \right]
\end{equation}
where
\begin{equation} 
\eta = \frac{NKT}{2\, \nu^{(r)}}
\end{equation}

is the dynamic viscosity. Since $\nu^{(r)}$ is proportional to $N$, $\eta$ is independent of $N$. The viscosity given by (38) agrees quantitatively with the corresponding result provided by the first approximation of the Chapman--Enskog theory (see Ref.~1), thereby  showing that a properly designed relaxation model has more to offer than bare simplicity.

Next we have to derive a corresponding relation for {\bf q}. This turns out to be a highly intricate matter, which may appear surprising  in view of the fact that the anticipated result, ${\bf q} = - \kappa\, \nabla T$ (wih $\kappa $ the heat conductivity), is as simple as a result can be. In fact, {\bf q} depends on a wealth of terms, and it will be necessary to find conditions allowing to ignore the majority of them. One complication will be that T4 in this case cannot be ignored, so that moments of order $n=4$ will come into play. 

We consider the $x$--component of {\bf q} which, according to (17), requires knowledge of the three moments $M^{(3,0,0)}$, $M^{(1,2,0)}$ and $M^{(1,0,2)}$. From (7)--(10):

For $(i,j,k) = (3,0,0)$

\begin{equation} 
{\rm T1} = \frac{dM^{(3,0,0)}}{dt} + M^{(3,0,0)}\Big( \,\nabla\cdot {\bf v} + 3\,\frac{\partial v_x}{\partial x} \Big)  - \frac{\delta M^{(3,0,0)}}{\delta t}
\end{equation}

\begin{equation} 
{\rm T2} =  3\, p_{xx}\,\Big(\frac{dv_x}{dt}-\frac{F_x}{m} \Big) 
\end{equation}

\begin{equation} 
{\rm T3} =  3\, \Big(M^{(2,1,0)}\,\frac{\partial v_x}{\partial y} + \,M^{(2,0,1)}\,\frac{\partial v_x}{\partial z} \Big) 
\end{equation} 

\vspace*{-2mm}
\begin{equation} 
{\rm T4} = \frac{\partial M^{(4,0,0)}}{\partial x} + \frac{\partial M^{(3,1,0)}}{\partial y} + \frac{\partial M^{(3,0,1)}}{\partial z}
\end{equation}

For $(i,j,k) = (1,2,0)$

\begin{equation} 
{\rm T1} = \frac{dM^{(1,2,0)}}{dt} + M^{(1,2,0)}\Big( \,\nabla\cdot {\bf v} + \frac{\partial v_x}{\partial x} + 2\, \frac{\partial v_y}{\partial y} \Big)  - \frac{\delta M^{(1,2,0)}}{\delta t}
\end{equation}

\begin{equation} 
{\rm T2} =  p_{yy} \Big(\frac{dv_x}{dt}-\frac{F_x}{m} \Big) + 2\, p_{xy} \Big(\frac{dv_y}{dt}-\frac{F_y}{m} \Big) 
\end{equation}

\begin{equation} 
{\rm T3} =  \Big( M^{(0,3,0)}\,\frac{\partial v_x}{\partial y} + M^{(0,2,1)}\,\frac{\partial v_x}{\partial z} \Big) +
2\, \Big( M^{(2,1,0)}\,\frac{\partial v_y}{\partial x} + M^{(1,1,1)}\,\frac{\partial v_y}{\partial z} \Big) 
\end{equation} 

\vspace*{-2mm}
\begin{equation} 
{\rm T4} = \frac{\partial M^{(2,2,0)}}{\partial x} + \frac{\partial M^{(1,3,0)}}{\partial y} + \frac{\partial M^{(1,2,1)}}{\partial z}
\end{equation}

For $(i,j,k) = (1,0,2)$

\begin{equation} 
{\rm T1} = \frac{dM^{(1,0,2)}}{dt} + M^{(1,0,2)}\Big( \,\nabla\cdot {\bf v} + \frac{\partial v_x}{\partial x} + 2\, \frac{\partial v_z}{\partial z} \Big)  - \frac{\delta M^{(1,0,2)}}{\delta t}
\end{equation}

\begin{equation} 
{\rm T2} =  p_{zz} \Big(\frac{dv_x}{dt}-\frac{F_x}{m} \Big) + 2\, p_{xz} \Big(\frac{dv_z}{dt}-\frac{F_z}{m} \Big) 
\end{equation}

\begin{equation} 
{\rm T3} =  \Big( M^{(0,1,2)}\,\frac{\partial v_x}{\partial y} + M^{(0,0,3)}\,\frac{\partial v_x}{\partial z} \Big) +
2\, \Big( M^{(2,0,1)}\,\frac{\partial v_z}{\partial x} + M^{(1,1,1)}\,\frac{\partial v_z}{\partial y} \Big) 
\end{equation} 

\vspace*{-2mm}
\begin{equation} 
{\rm T4} = \frac{\partial M^{(2,0,2)}}{\partial x} + \frac{\partial M^{(1,1,2)}}{\partial y} + \frac{\partial M^{(1,0,3)}}{\partial z}
\end{equation}

Our first approximative assumption will be that the collision term in the three expressions for T1 dominates the other terms, with the consequence that the collision term will be the only one to survive. By the same token, T3 vanishes altogether. 
Consequently, if we add the three T1 and T3 terms (the sum denoted by the $\Sigma$--sign), we obtain the simple result

\begin{equation} 
\Sigma T1 + \Sigma T3 = - 2\,\frac{\delta q_x}{\delta t}
\end{equation} 

For the treatment of the T2--terms we use condition (31) to neglect the $p_{ij}$--terms and replace the $p_{ii}$--terms by $p$, giving

\begin{equation} 
\Sigma T2 = 5\, p \,\Big(\frac{dv_x}{dt}-\frac{F_x}{m} \Big)
\end{equation}

The terms T4, as they stand, open the path to an endless number of equations for moments of order $n \ge 4$. To avoid this hopeless endeavour, we proceed as follows: We divide the moments of order $n=4$ into even and odd moments, $M_{even}^{(i,j,k)}$ and $M_{odd}^{(i,j,k)}$, where even means that all indices $i$, $j$, $k$ are even (0 regarded as an even number), and odd that this is not the case. Further we introduce the moments $M_{maxw}^{(i,j,k)}$ which are obtained by use of the Maxwellian (2). These definitions are the basis for the conditions 

\begin{equation} 
|M_{odd}^{(i,j,k)}| \ll M_{even}^{(i,j,k)} \mbox{\hspace{.6cm} and \hspace{.6cm}} |M_{even}^{(i,j,k)} - M_{maxw}^{(i,j,k)}| \ll M_{maxw}^{(i,j,k)}
\end{equation}

which is similar to condition (31). The background for the conditions (53) is that odd moments occur only in the form of a deviation from equilibrium, whereas even moments also exist in equilibrium. Imposing (53) on the terms T4, we obtain

\begin{equation} 
\Sigma T4 = \frac{\partial}{\partial x} \left( M^{(4,0,0)}_{\it maxw} + M^{(2,2,0)}_{\it maxw} + M^{(2,0,2)}_{\it maxw} \right)
\end{equation} 

Putting the pieces together, we arrive at

\begin{equation} 
\frac{\delta q_x}{\delta t} = \frac{5}{2}\,p \,\Big(\frac{dv_x}{dt}-\frac{F_x}{m} \Big)
+ \frac{1}{2}\, \frac{\partial}{\partial x} \left( M^{(4,0,0)}_{\it maxw} + M^{(2,2,0)}_{\it maxw} + M^{(2,0,2)}_{\it maxw} \right)
\end{equation}

With the useful relation

\begin{equation} 
M^{(i,j,k)}_{\it maxw} = (i-1)!!\, (j-1)!!\, (k-1)!!\,\, NKT\, V^{n-2} \, \delta^{(i,j,k)}
\end{equation}

(where $\delta^{(i,j,k)} = 1$ if $i$, $j$, $k$ are all even ; $\delta^{(i,j,k)} = 0$ otherwise), the terms in the second brackets in (55) add up to $5NK^2T^2/m $. Then, approximating $(dv_x/dt - F_x/m)$ by the leading term in (24), $-(1/Nm)\, \partial(NKT)/\partial x$, and generalizing from $\delta q_x/\delta t$ to $\delta {\bf q}/\delta t$, we obtain without further approximation

\begin{equation} 
\frac{\delta {\bf q}}{\delta t} = \frac{5}{2} NKV^2 \, \nabla \, T 
\end{equation}

Treating the relation between $\delta {\bf q}/ \delta t$ and {\bf q} in the same way as we did for $\delta (\textsf{p})^{\rm o} / \delta t$ and $(\textsf{p})^{\rm o}$, we obtain the final result

\begin{equation} 
{\bf q} = - \, \kappa \, \nabla \, T
\end{equation} 

with the heat conductivity $\kappa$ given by

\begin{equation} 
\kappa = \frac{5\,NK\,V^2}{4\, \nu^{(r)}}
\end{equation}

We note in passing that the result we would have obtained by neglecting moments of order $n=4$ would be the same in magnitude as (58), but with opposite sign, i.e., heat would flow at the correct rate, but in the wrong direction from cold to hot. 

Whereas the viscosity coeffient $\eta$ given by (38) agreed fully with the first approximation of the Chapman--Enskog theory, (59) should be corrected by replacing the factor 5/4 by 15/8 to attain corresponding agreement for $\kappa$. The reason for the discrepancy is that the collision model used (Ref.~6) employs a relaxation function which involves 10 moments of $f$ (the first ten moments up to the elements of the pressure tensor), where it would have to involve 13 moments in order to be fully suited for a quantitative reproduction of the heat conductivity.

After having seen the full complexity of the equations (39)--(50), it appears like a miracle that such a simple result, eq.~(58), could  have been obtained in the end. But how is it possible that a result as plausible as (58) should require such a complicated derivation? Is it not obvious that heat will flow from hot to cold, at a rate proportional to the temperature gradient? In fact, the result (58) is deceptive by concealing that {\bf q} actually depends on a wealth of terms which all have to be negligibly small for (58) to be valid.
The importance of our treatment is not that we have recovered the well--known result (58), and likewise (37), but that the applicabilty limits of these equations have been worked out thoroughly.

We have now reached our goal to obtain expressions for $(\textsf{p})^{\rm o}$ and {\bf q} depending only on the fluid variables $N$, {\bf v} and $T$, thereby converting the open--ended system [(23),(26),(28)] into a closed system.

\vspace*{5mm} 
{\bf \large 6. The Navier--Stokes system}

Insertion of (37) and (58) in (26) and (28), treating $\eta$ and $\kappa$ as constants, yields 
\begin{equation} 
\frac{d \bf v}{d t} =   \frac{{\bf F}}{m} - \,  V^2 \left[\, \frac{\nabla T}{T} + \frac{\nabla N}{N} \, \right] +\frac{\eta}{N m} \, \nabla \cdot  \left[\, \nabla {\bf v} + (\nabla {\bf v})^t - \frac{2}{3}\, (\nabla\cdot {\bf v}) \, \textsf{U} \, \right]
\end{equation}

\begin{equation} 
\frac{dT}{dt} = -\frac{2}{3}\, T \, \nabla \cdot {\bf v}
+ \frac{2}{3}\, \frac{\eta}{KN}\, \left[ \nabla {\bf v} : [\nabla {\bf v} + (\nabla {\bf v})^t] - \frac{2}{3} (\nabla\cdot {\bf v})^2 \right] + \frac{2}{3}\, \frac{\kappa}{KN}\, \Delta T
\end{equation}
 
The system [(23),(60),(61)] may be called the Navier--Stokes system, since eq.~(60), corresponding to the Navier--Stokes equation, is the central element in the system. 

It may be irritating to see that the Navier--Stokes equation (60) contains only one viscosity coefficient, while at other places in the literature one finds that two different viscosity coefficients are needed. As an example, eq.~(15.6) of Landau and Lifshitz$^8$, written here in a form appropriate for easy comparison with (60), reads 

\begin{equation} 
\frac{d \bf v}{d t} =   \frac{{\bf F}}{m} - \,  V^2 \left[\, \frac{\nabla T}{T} + \frac{\nabla N}{N} \, \right] +\frac{\eta}{N m} \, \nabla \cdot  \left[\, \nabla {\bf v} + (\nabla {\bf v})^t - \frac{2}{3}\, (\nabla\cdot {\bf v}) \, \textsf{U} \, \right]
+ \frac{\zeta}{Nm}\, \nabla \cdot (\nabla \cdot {\bf v})\, \textsf{U}
\end{equation}

Here, $\zeta$ is the so--called volume viscosity. The question is now: Which of the two Navier--Stokes equations, (60) or (62), is the right one. The answer is: In the fluid limit (20), $\zeta$ vanishes, and hence (60) is the correct form of the Navier--Stokes equation. This statement is justified by the treatment in Ref.~4: 

In Ref.~4, a momentum equation (eq.~(91)) is derived, having the appearance of the Navier--Stokes equation, however with transport coefficients valid both inside and outside of the fluid regime. For a comparison of eq.~(62) with eq.~(91) of Ref.~4, we write the viscosity terms in (62) in the equivalent form

\begin{equation} 
\eta \, \nabla \cdot  \left[\, \nabla {\bf v} + (\nabla {\bf v})^t - \frac{2}{3}\, (\nabla\cdot {\bf v}) \, \textsf{U} \, \right] + \zeta\, \nabla \cdot (\nabla \cdot {\bf v})\, \textsf{U}= \frac{4}{3}\,\eta'\,\nabla (\nabla \cdot {\bf v}) - \eta\,\nabla\times(\nabla\times{\bf v})
\end{equation}
with
\begin{equation} 
\eta' = \eta + \frac{3}{4}\,\zeta \mbox{\hspace*{7mm} {\rm or} \hspace*{7mm}} \zeta = \frac{4}{3}\,(\eta'-\eta)
\end{equation}

General expressions for $\eta$ and $\eta'$ are given by eqs.~(56) and (93) of Ref.~4. Following the discussion there on page 31, the  conclusions are: In the general fluid limit (20), $\eta'$ approaches $\eta$, with the consequence that $\zeta$ vanishes, and in the collisional fluid limit (22), $\eta'$ and $\eta$ furthermore become matter constants. Thus, (60) is indeed the correct version of the Navier--Stokes equation, provided it is applied inside the range set by condition (22). Outside this range, a second viscosity builds up, however not as a matter constant, but as a coefficient that depends both on the temporal and spatial evolution of the considered process. 

It is a common misbelief that the Navier--Stokes system relates to all fluids, from ideal gases via real gases to liquids, differing only with respect to the numerical values of the transport coefficients $\eta$ and $\kappa$. However, the system [(23),(60),(61)] contains no terms to account for the cumulative action of intermolecular forces and for the finite size of the particles, and thus it relates to ideal gases. We will come back to this in Section 10.

\vspace*{5mm} 
{\bf \large 7. Energy conservation}
 
The system [(23),(60),(61)] must be energy conservative. In order to check this, we convert (60) and (61) into energy equations: 

An ensemble of particles with individual velocities {\bf u} has the energy density

\begin{equation} 
w =  \frac{1}{2}\, m \int \! f \, {\bf u}^2 d^3 {\bf u} = \frac{1}{2}\, m \,\bigg( \int \! f \, {\bf v}^2 d^3 {\bf u} + \int \! f \, ({\bf u}-{\bf v})^2 d^3 {\bf u} + 2 \! \int \! f \, {\bf v}\,({\bf u} - {\bf v}) \, d^3 {\bf u} \bigg)  
\end{equation}

The first RHS term represents the kinetic energy density due to the bulk motion,

\begin{equation} 
w_K =  \frac{1}{2}\, m \int f \, {\bf v}^2 d^3 {\bf u} =  \frac{1}{2}N m{\bf v}^2
\end{equation}

the second the internal energy density,

\begin{equation} 
w_I = \frac{1}{2}\, m \int f \, ({\bf u} -{\bf v})^2 d^3{\bf u}  = \frac{3}{2}\, NKT  
\end{equation}

and the third vanishes. 

From (24) (multiplied by {\bf v} and with {\bf F} set to zero) and (28), both in conjunction with the continuity equation (23), the energy equations

\begin{equation} 
\frac{\partial w_K}{\partial t} + \nabla \cdot ( w_K \, {\bf v}) = - \, {\bf v}\cdot \left[\, \nabla p + \nabla \cdot (\textsf{p})^{\rm o} \, \right]
\end{equation}

\begin{equation} 
\frac{\partial w_I}{\partial t} + \nabla \cdot (w_I \, {\bf v}) = - \, \left[\, p \, \nabla \cdot {\bf v} +(\textsf{p})^{\rm o} : \nabla {\bf v} \right]  - \nabla \cdot {\bf q}
\end{equation}

are obtained. With the identity $\; {\bf v}\cdot [\,  \nabla \cdot (\textsf{p})^{\rm o}] + (\textsf{p})^{\rm o} : \nabla {\bf v} = \nabla\cdot [ (\textsf{p})^{\rm o}\cdot {\bf v}]  $ , the sum of $w_K$ and $w_I$, $w = w_K + w_I$, is given by

\begin{equation} 
\frac{\partial w}{\partial t} + \nabla \cdot (w \, {\bf v}) = - \nabla \cdot (p \, {\bf v}) -  \nabla\cdot [ (\textsf{p})^{\rm o}\cdot {\bf v}]  - \nabla \cdot {\bf q}
\end{equation}

Insertion of (37) and (58) in (68)--(70) yields

\begin{equation} 
\frac{\partial w_K}{\partial t} + \nabla \cdot ( w_K \, {\bf v}) = - \, {\bf v}\cdot \nabla p +  \eta \, {\bf v}\cdot  \left( \nabla \cdot \left[\, \nabla {\bf v} + (\nabla {\bf v})^t - \frac{2}{3}\, (\nabla\cdot {\bf v}) \, \textsf{U} \, \right] \right)
\end{equation}

\begin{equation} 
\frac{\partial w_I}{\partial t} + \nabla \cdot (w_I \, {\bf v}) = - \,   p \, \nabla \cdot {\bf v} + \eta \, \left[ \nabla {\bf v} : [\nabla {\bf v} + (\nabla {\bf v})^t] - \frac{2}{3} (\nabla\cdot {\bf v})^2 \right] + \kappa \, \Delta T
\end{equation}

\begin{equation} 
\frac{\partial w}{\partial t} + \nabla \cdot (w \, {\bf v}) = - \nabla \cdot (p \, {\bf v}) +  \eta \, \nabla \cdot  \left( {\bf v} \cdot \left[\, \nabla {\bf v} + (\nabla {\bf v})^t - \frac{2}{3}\, (\nabla\cdot {\bf v}) \, \textsf{U} \, \right] \right) + \kappa \, \Delta T
\end{equation}

With Gau\ss' integral theorem, a consequence of (70) or (73) is that the total energy  $E = \int \! w \, d^3 {\bf r}$ inside a closed solid surface, or in entire space, is constant,

\begin{equation} 
\frac{dE}{dt} = 0
\end{equation}

and this proves that the Navier--Stokes system in its complete form [(23),(60),(61)] is indeed energy conservative. Naturally, the conservation of energy relates to $E$ in total, not to its parts $E_K = \int \! w_K \, d^3 {\bf r}$ and $E_I = \int \! w_I \, d^3 {\bf r}$ individually. Thus, a violation of the integrity of the complete Navier--Stokes system would have the consequence that energy is not conserved.

If we express the right--hand sides of (71) and (72) in component form, it will be difficult to attach a physical meaning to all the terms arising. We can, however, bundle terms in a useful way and get a better physical understanding thereby. We do this by introducing the following abbreviations:

\begin{equation} 
\dot{w}_{K\leftrightarrow P} = - \, {\bf v} \cdot \nabla p + 2 \, \eta \left[ v_x \frac{\partial^2 v_x}{\partial x^2} +  v_y \frac{\partial^2 v_y}{\partial y^2} +  v_z \frac{\partial^2 v_z}{\partial z^2} \, \right] - \frac{2}{3} \eta \, \, {\bf v} \cdot \nabla ( \nabla \cdot {\bf v})
\end{equation}

\begin{equation} 
\dot{w}_{I\leftrightarrow P} = - \, p \, \nabla \cdot {\bf v} + 2 \, \eta \bigg[ \left( \frac{\partial v_x}{\partial x} \right)^{\mbox{\hspace*{-1.2mm}}2} + \left( \frac{\partial v_y}{\partial y} \right)^{\mbox{\hspace*{-1.2mm}}2} + \left( \frac{\partial v_z}{\partial z} \right)^{\mbox{\hspace*{-1.2mm}}2} \, \bigg] - \frac{2}{3} \eta \, \, (\nabla \cdot {\bf v} )^2
\end{equation}

\begin{equation} 
\dot{w}_{K\leftrightarrow K} = \eta \left\{ \frac{\partial}{\partial x} \left[ v_y \left( \frac{\partial v_x}{\partial y} + \frac{\partial v_y}{\partial x} \right) + v_z \left( \frac{\partial v_x}{\partial z} + \frac{\partial v_z}{\partial x} \right) \right] 
+ ..... + ..... \, \, \right\}
\end{equation}

\begin{equation} 
\dot{w}_{I\leftrightarrow I} = \kappa \, \nabla\cdot (\nabla T) 
\end{equation}

\begin{equation} 
\dot{w}_{K\rightarrow I} = \eta \, \bigg[ \left( \frac{\partial v_x}{ \partial y} 
+ \frac{\partial v_y}{\partial x} \right)^{\mbox{\hspace*{-1.2mm}}2} + \left( \frac{\partial v_y}{\partial z} + \frac{\partial v_z}{\partial y} \right)^{\mbox{\hspace*{-1.2mm}}2} +\left( \frac{\partial v_z}{\partial x} + \frac{\partial v_x}{\partial z} \right)^{\mbox{\hspace*{-1.2mm}}2} \, \bigg]
\end{equation}

whereby the energy equations (71) and (72) adopt the short form  

\begin{equation} 
\frac{\partial w_K}{\partial t} + \nabla \cdot ( w_K
 \, {\bf v})  =  \dot{w}_{K\leftrightarrow P} + \dot{w}_{K\leftrightarrow K} - \dot{w}_{K\rightarrow I}
\end{equation}

\begin{equation} 
\frac{\partial w_I}{\partial t} + \nabla \cdot (w_I \, {\bf v}) =
\dot{w}_{I\leftrightarrow P} + \dot{w}_{I\leftrightarrow I} +
\dot{w}_{K\rightarrow I}
\end{equation}

The energy transfer rates on the right--hand sides of (80) and (81) have the following physical meanings, as indicated by the chosen nomenclature: $\dot{w}_{K\leftrightarrow P}$ describes the mutual conversion of kinetic and potential energy of an ensemble of particles moving along or opposite to  the pressure gradient, $\dot{w}_{I\leftrightarrow P}$ descibes the mutual conversion of internal and potential energy due to compression or dilatation, $\dot{w}_{K\leftrightarrow K}$ describes the spatial redistribution of kinetic energy under the action of viscosity, $\dot{w}_{I\leftrightarrow I}$ describes the spatial redistribution of internal energy due to conduction of heat, and $\dot{w}_{K\rightarrow I}$, being unconditionally positive, describes the one--sided conversion of kinetic into internal energy (viscous heating).

Integration over a volume enclosed by a solid surface yields:  

\begin{equation} 
\frac{dE_{K\leftrightarrow K}}{dt} \equiv \int \! \dot{w}_{K\leftrightarrow K} \, d^3{\bf r} = 0
\end{equation}
 
\begin{equation} 
\frac{dE_{I\leftrightarrow I}}{dt} \equiv \int \! \dot{w}_{I\leftrightarrow I} \, d^3{\bf r} = 0
\end{equation}

\begin{equation} 
\frac{dE_{K\leftrightarrow P}}{dt} + \frac{dE_{I\leftrightarrow P}}{dt} \equiv \int (\dot{w}_{K\leftrightarrow P} + \dot{w}_{I\leftrightarrow P}) \, d^3 {\bf r} = 0
\end{equation}

Note that only the sum of $dE_{K\leftrightarrow P}/dt$ and $dE_{I\leftrightarrow P}/dt$ is zero, not these terms alone. Further
 
\begin{equation} 
\frac{dE_{K\rightarrow I}}{dt} \equiv \int \! \dot{w}_{K\rightarrow I} \, d^3{\bf r} > 0
\end{equation}

From (80) to (84) the rate of change of the total energy within the given volume follows as

\begin{equation} 
\frac{dE}{dt} = \underbrace{ \frac{dE_{K\leftrightarrow K}}{dt} }_{= \, 0} + \underbrace{  \frac{dE_{I\leftrightarrow I}}{dt} }_{= \, 0} + \underbrace{ \frac{dE_{K\leftrightarrow P}}{dt} + \frac{dE_{I\leftrightarrow P}}{dt} }_{= \, 0} 
+ \underbrace{ \frac{dE_{K\rightarrow I}}{dt} - \frac{dE_{K\rightarrow I}}{dt} }_{= \, 0} = 0
\end{equation} 

and this is another way to show that the Navier--Stokes system in its complete form [(23),(60),(61)] is energy conservative.

\vspace*{5mm} {\bf \large 8. Applicability conditions summarized}

Fluid theory in general, and the Navier--Stokes system in particular, are bound to a number of applicability conditions which have been given above, scattered over the text. These will be summarized as follows  :

1. In the treatment above, leading to the Navier--Stokes system [(23),(60),(61)], particles are treated as points without finite size and without intermolecular forces. The applicability of this system is thus restricted to ideal gases. In order to set a limit separating ideal gases from real fluids, we employ van der Waals' equation which provides a relation $p(N,T)$ for real fluids:

\begin{equation} 
p = \frac{NKT}{1- V_0N} - \frac{a}{A^2}N^2
\end{equation}

Here, $A$ is Avogadro's number, $V_0$ is a measure of the volume occupied by a single particle, and $a$ is a tabulated quantity determining the strength of the resultant intermolecular force, averaged over a large number of neighbouring particles. The pressure given by (87) is a measure of the total force exerted on a surface by a real fluid, whereas the pressure defined by (15), which is contained in (87), describes the transfer of momentum due to thermal motion.

Eq.~(87) shows that the ideal gas state is characterized by the conditions

\begin{equation} 
N \ll \frac{1}{V_0} \mbox{\hspace*{8mm} and \hspace*{8mm}} N \ll \frac{KTA^2}{a}
\end{equation}

A fluid not obeying theses conditions is not suitable for application of the Navier--Stokes system [(23),(60),(61)].

2. To proceed from the infinite system of moment equations (6)--(10) to a finite set of fluid equations, condition (20) has to be satisfied. In the context of the present paper, it is the collisional limit of (20), 
\begin{equation}\tag{22} 
\tau_c \ll \tau \mbox{\hspace*{8mm} {\rm and} \hspace*{8mm}} \tau_c V \ll l
\end{equation}

which has to be applied. This condition secures that different moments can be connected via local transport coefficients (e.g., ${\bf q} = - \kappa\, \nabla T$). `Local' implies that the transport coefficients in question are matter constants, i.e., independent of the characteristics of the process. A generalization from local to nonlocal is given in Ref.~4.

A second, but equivalent function of condition (22) is to justify the dominance of the collision term over other terms. This is seen most clearly in the derivation of the result ${\bf q} = - \kappa\, \nabla T$ where all terms except the collision term have been neglected in T1 and T3 (see (51)). 

3. A condition of equal importance concerns the different weights of even and odd moments. A moment is called even if all indices ($i,j,k$) are even (zero taken as an even number), and odd otherwise. The corresponding condition, given above by eq.~(53), reads

\begin{equation}\tag{53} 
|M_{odd}^{(i,j,k)}| \ll M_{even}^{(i,j,k)} \mbox{\hspace{.8cm} and \hspace{.8cm}} |M_{even}^{(i,j,k)} - M_{maxw}^{(i,j,k)}| \ll M_{maxw}^{(i,j,k)}
\end{equation}

where $M_{maxw}^{(i,j,k)}$ corresponds to $M_{even}^{(i,j,k)}$ if the latter is obtained by using a Maxwell distribution. Condition (31), formulated for the elements of the pressure tensor, is physically equivalent to (53). 

The physical background for condition (53) is that odd moments occur only in the form of a deviation from equilibrium, whereas even moments also exist in equilibrium. Thus, (53) rules out solutions far away from equilibrium and thereby has the character of a hidden linearization. Since condition (53) is applied throughout the derivation of the Navier--Stokes system, it is an absolute necessity that solutions of the Navier--Stokes system obey (53) in order to be regarded as physically valid. One possibilty to check this a posteriori is to calculate $p_{ii}$ and $p_{ij}$ following from (37) and ensure that $|p_{ij}| \ll p_{ii}$ and $|p_{ii} - p| \ll p$.

Condition (53) is a strong constraint for the applicabiliy of the Navier--Stokes system [(23),(60),(61)], but without it, it would not be possible to get a closed system of equations, decoupled from higher--order moments. One consequence of this condition is that fluid equations can be extracted only pairwise from the infinite set of moment equations, where the first of these has to describe an odd moment, and the second an even moment. In our case, the pair consists of eqs.~(60) and (61).  
It is important to note that this pair is inseparable. If it is separated nonetheless, as in certain simplified versions of the Navier--Stokes system, the treatment becomes inconsistent, insofar as terms of given order are kept in one part of the system, but disregarded in the other.

The above conditions [(88),(22),(53)] set a tight margin to the physical validity of the Navier--Stokes system [(23),(60),(61)]. It will be easily possible to get into solutions which are mathematically correct, but physically invalid. A validity check a posteriori will therefore be a need.

\newpage
{\bf \large 9. The assumption of incompressibility}

The assumption of incompressibility is found in a vast number of publications, used with the intention to make the Navier--Stokes system simpler. Incompressibility means that the density of a fluid parcel cannot be changed. However, as we have seen, the Navier--Stokes system [(23),(60),(61)] contains no term that would prevent compressing a fluid parcel to an arbitrarily small volume.  Thus, incompressibility here is not meant as a material property, but as the property of a process. Thereby, the assumption of incompressibility acts like a filter, sorting out a great number of physical processes and concentrating on those which hopefully allow the desired simplification. 

Under the assumption of incompressibility, the continuity equation (23) is replaced by 

\begin{equation} 
N = {\rm const.} \,\, \Longrightarrow \,\, \nabla \cdot {\bf v} = 0
\end{equation}

The assumption of incompressibility is commonly justified by the Mach condition

\begin{equation} 
|{\bf v}| \, \ll \, V_s \, \approx \, V
\end{equation}

where $V_s$ is the sound velocity. The Mach condition, important as it may be for aeronautics purposes, has no impact on the the Navier--Stokes system. Following the derivation in Section 5 step by step, one sees that there exists no place where condition (90) could be implemented. Thus, the Navier--Stokes system [(23),(60),(61)] is the same without or with the Mach condition. 

All one can do to make use of (89) is to externally impose it on [(23),(60),(61)], giving:

\begin{equation} 
\frac{d \bf v}{d t} = \frac{{\bf F}}{m} - \,  V^2 \, \frac{\nabla T}{T}\, + \, \frac{\eta}{N m} \, \nabla \cdot  [\, \nabla {\bf v} + (\nabla {\bf v})^t ]
\end{equation}

\begin{equation} 
\frac{dT}{dt} =  \frac{2}{3}\, \frac{\eta}{KN}\,  \nabla {\bf v} : [\nabla {\bf v} + (\nabla {\bf v})^t]  + \frac{2}{3}\, \frac{\kappa}{KN}\, \Delta T   
\end{equation}

Thereby, the closed 3--equation system [(23),(60),(61)] is reduced to the closed 2--equation system [(91),(92)], and the continuity equation in the form (89) is no longer needed after it has done its duty to simplify (60) and (61). In (91), the temperature gradient $\nabla T$ is left as the only driving agent after $\nabla N$ has been removed due to (89). 

Certainly, the system [(91),(92)] is simpler than the complete system [(23),(60),(61)], but still too complicated for an analytical treatment, and hence rather pointless. Moreover, it has to be realized that the system [(91),(92)] will most likely not produce a  velocity field obeying $\nabla \cdot {\bf v} = 0$, and thereby be in contradiction with the starting condition. Altogether, the system [(91),(92)] is seen to be useless.

It is worth noting that in stringent treatises on transport theory, based on kinetic theory rather than heuristic assumptions, the term `incompressibility' does not exist at all (e.g., Refs.~1 and 3). A fluid behaves as it behaves, not as one may want it to behave. Indeed, a valid theory is complete and determines self--consistently to what extent the fluid might behave as if incompressible. The Navier--Stokes system in the form [(23),(60),(61)] is complete in this sense. 

This could be the end of our discussion on incompressibility. However, there exists another even simpler version of a Navier--Stokes system in the literature, primarily in the applied mathematics literature, namely

\begin{equation} 
\frac{d \bf v}{d t} =  \frac{\bf F}{m} - \,  V^2 \, \frac{ \nabla T}{T}\, +\, \frac{\eta}{N m} \, \,\Delta {\bf v}
\end{equation}

\begin{equation} 
\nabla \cdot {\bf v} = 0
\end{equation}

Eq.~(93) follows from (91) with the identity $\nabla \cdot  [\, \nabla {\bf v} + (\nabla {\bf v})^t] = \Delta {\bf v} + \nabla ( \nabla \cdot {\bf v} )$, and (94) supercedes the thermal equation (92). 

The system [(93),(94)] represents the so--called ``incompressible Navier--Stokes equations''. A prominent example for its application  is the Millennium Prize, endowed by the Clay Mathematics Institute (Fefferman$^9$) with the ambitious expectation to unlock the secrets hidden in the Navier--Stokes equations. This appears to be a rather optimistic aim, considering that the system [(93),(94)] has little in common with the complete Navier--Stokes system [(23),(60),(61)].

Despite its frequent use in the literature, the system [(93),(94)] must be rejected since it is physically deficient in several ways:
 
1. Eq.~(93) calls for a closing equation for the temperature $T$. It is obvious that this equation would have to comprise the basic thermal processes, i.e., heat transport by conduction and convection, and heat generation by viscous heating. The closing equation (94), used in place of an appropriate thermal equation, is entirely useless in this regard.

2. It has been seen in connection with condition (53) that the thermal equation is an irremovable part of the Navier--Stokes system.  A neglect of the thermal equation leads to the inconsistency that terms of given order of magnitude are kept in one part of the system, but disregarded in the other.

3. In the practical handling of [(93),(94)], the role assigned to $T({\bf r},t)$ is to ensure that the resulting ${\bf v}({\bf r},t)$ satisfies the incompressibility condition (94). This means that $T$ is treated as a dummy, deprived of its thermodynamic properties. But it is not a matter of free choice to assign a special role to $T$. The temperature is what it is, a thermodynamic quantity, determined by the thermal equation, and this equation does not lose its existence only because incompressibility has been assumed. Thus, any solution of [(93),(94)], if intended to be physically valid, will have to be consistent with the thermal equation, even though the latter is not incorporated in the system. So, in effect, the thermal equation is back and reoccupies its place, eliminating the same equation $\nabla \cdot {\bf v} = 0$ by which it had been eliminated before.

4. A decisive question is whether the system [(93),(94)] conserves the total energy. The absence of a thermal equation has the consequence that the internal energy equation (81) vanishes, whereby eq.~(86) is altered to

\begin{equation} 
\frac{dE}{dt} = \underbrace{ \frac{dE_{K\leftrightarrow K}}{dt} }_{= \, 0} + \underbrace{ \frac{dE_{K\leftrightarrow P}}{dt}}_{\neq \, 0} - \underbrace{\frac{dE_{K\rightarrow I}}{dt}}_{> \, 0} \neq 0
\end{equation} 

showing that the total energy is not conserved. This is an absolute disqualifier for the ``incompressible Navier--Stokes equations'' [(93),(94)].

As a consequence of these four points, work based on the ``incompressible Navier--Stokes equations'' has to be understood as mathematical work outside the range of justified physics.

\newpage
{\bf \large 10. Extension to real fluids}

As noted before, the Navier--Stokes system [(23),(60),(61)] is not applicable to dense gases and liquids. An extension of these equations will have to take into account the finite size of particles and the intermolecular forces between them. It appears unlikely that this could be done in a rigorous fashion. This will limit our ambitions and suggest that we should proceed in a pragmatic way. We will do this by taking recourse to van der Waals' equation in combination with the treatment in Chapter 16 of Chapman and Cowling$^1$. 

Van der Waals' equation, given by (87), is a classical example for pragmatic physics. Being without stringent derivation, van der Waals' equation ows its justification to its fair agreement with experimental results (outside the so--called Maxwell area), covering the range from ideal gases via dense gases (before reaching the Maxwell area) to liquids (after leaving the Maxwell area). In particular, the question of compressibe vs.~incompressibile is self--consistently settled by (87): A compression beyond $N V_0 = 1$ is not possible. 

The pressure gradient following from (87) reads

\begin{equation} 
\nabla p = \nabla (NKT) + \nabla \left( \frac{V_0N}{1-V_0N}\, NKT  \right) - \frac{2a}{A^2}\, N \nabla N  
\end{equation}
or
\begin{equation} 
\nabla p = NmV^2 \left[\, \varphi \frac{\nabla T}{T} + \varphi^2 \frac{\nabla N}{N} \, \right] - \frac{2a}{A^2}\, N \nabla N  
\end{equation}
with
\begin{equation} 
\varphi(N) = \frac{1}{1-V_0N}
\end{equation}

The first RHS term in (96) relates to a gas consisting of point particles without intermolecular forces, the second is a correction to account for the finite size of particles, and the third describes the accumulated action of intermolecular forces. 

The question is now how to use (96) in order to modify the system [(23),(26),(28)] in the intended way. Regarding the continuity equation (23),  a change is neither needed nor possible: The continuity equation counts the difference between the number of particles entering and leaving a volume element. This balance is independent of the size of particles and of forces between them.

Next turning to the question of how to incorporate intermolecular forces: A test particle is subject to forces from all surrounding particles. In a homogeneous fluid, these forces average out to zero, but in an inhomogeneous fluid a net force remains that pulls the particle from lower to higher densities. This is expressed by the last term in (96). For inclusion in (26), the force {\bf F} must be replaced by

\begin{equation} 
{\bf F} = {\bf F}_{ext} + {\bf F}_{int} \mbox{\hspace*{8mm} {\rm with} \hspace*{8mm}} {\bf F}_{int} =   \frac{2a}{A^2}\, \nabla N
\end{equation}

where ${\bf F}_{ext} $ is what {\bf F} has been before, viz.~an external force acting on the particles individually.

What remains is to modify the terms $ \nabla T/T $ and $ \nabla N/N $ in (26), and $ T \, \nabla \cdot {\bf v} $ in (28). A comprehensive derivation, however restricted to the limit  $NV_0 \ll 1$, is given in Ref.~1, leading to the results (16.33.3) and (16.33.4). Written in our nomenclature, these results correspond to the replacements 

\begin{equation} 
\nabla T/T \rightarrow (1+V_0N)(\nabla T/T) \mbox{\hspace{8mm} and \hspace{8mm}}  \nabla N/N \rightarrow (1+2V_0N)(\nabla N/N)    
\end{equation}
and
\begin{equation} 
T \, \nabla \cdot {\bf v} \rightarrow (1+V_0N)\, T \, \nabla \cdot {\bf v}
\end{equation}

Now, since (100) and (101) are obtained for $NV_0 \ll 1$, we may write $(1+V_0N) = \varphi$ and $(1+2V_0N) = \varphi^2$. Insertion in (26) and (28) yields

\begin{equation} 
\frac{d\bf v}{dt} = \frac{1}{m}  ({\bf F}_{ext} + {\bf F}_{int}) -  V^2 \left[\,\varphi \frac{\nabla T}{T} + \varphi^2 \frac{\nabla N}{N} \, \right]  - \frac{1}{Nm} \nabla \cdot (\textsf{p})^{\rm o} 
\end{equation} 

\begin{equation} 
\frac{d T}{d t} = - \frac{2}{3} \,\varphi \, T \, \nabla \cdot {\bf v} - \frac{2}{3KN}\,\left[ \, (\textsf{p})^{\rm o} : \nabla {\bf v} +
\nabla \cdot {\bf q} \, \right]
\end{equation}

There exists formal agreement between the first RHS term in (97) and the second RHS term in (102). But whereas the latter is resricted to small $NV_0$, the first is understood as being valid for all $NV_0$. Thus, using van der Waals' equation for an extension of the Navier--Stokes system beyond ideal gases is equivalent to applying [(102),(103)] for all values of $NV_0$. There exists no theoretical justification for this step. It is the ability of van der Waals' equation to reproduce adequately experimental results on real gases and liquids that gives the justification, and this is the pragmatic aspect of our approach.   

The task left is to determine $(\textsf{p})^{\rm o}$ and {\bf q} for use in (102) and (103). In the kinetic equation (1), which is the mother equation for the whole derivation here, the first two terms apply to all fluids. The  distinction between ideal gases and real fluids has its source in the force term and the collision term. The derivation in Section 5 has shown that $(\textsf{p})^{\rm o}$ and {\bf q} do not depend on {\bf F}, and hence the results (35) and (57) remain unaltered. The problem is thus reduced to obtaining a connection between $\delta (\textsf{p})^{\rm o}/\delta t $ and $(\textsf{p})^{\rm o}$, and between $\delta {\bf q} / \delta t$ and ${\bf q} $. In the case of an ideal gas this problem has been solved quantitatively, resulting in (37) and (58), with the transport coefficients $\eta$ and $\kappa$ specified by (38) and (59). Here we will just assume that there exists a proportionality between $\delta (\textsf{p})^{\rm o}/\delta t $ and $(\textsf{p})^{\rm o}$, and between $\delta {\bf q} / \delta t$ and ${\bf q} $, without specifying the proportionality factors. Hereby, (37) and (58) remain applicable, however with the caveat that $\eta$ and $\kappa$ are now   empirical quantities without theoretical backing.  

Insertion of (37) and (58) in (102) and (103), respectively, yields

\begin{equation} 
\frac{d \bf v}{d t} =   \frac{1}{m} ({\bf F}_{ext} + {\bf F}_{int}) - \,  V^2 \left[\, \varphi \frac{\nabla T}{T} + \varphi^2 \frac{\nabla N}{N} \, \right] +\frac{\eta}{N m} \, \nabla \cdot  \left[\, \nabla {\bf v} + (\nabla {\bf v})^t - \frac{2}{3}\, (\nabla\cdot {\bf v}) \, \textsf{U} \, \right]
\end{equation}

\begin{equation} 
\frac{dT}{dt} = - \frac{2}{3} \,\varphi \, T \, \nabla \cdot {\bf v}
+ \frac{2}{3}\, \frac{\eta}{KN}\, \left[ \nabla {\bf v} : [\nabla {\bf v} + (\nabla {\bf v})^t] - \frac{2}{3} (\nabla\cdot {\bf v})^2 \right] + \frac{2}{3}\, \frac{\kappa}{KN}\, \Delta T
\end{equation}

The system [(23),(104),(105)] replaces the system [(23),(60),(61)], and they coincide for ${\bf F}_{int} = 0$ and $\varphi = 1$.

Since $\varphi > 1$, an important consequence of (104) is that density gradients have a higher weight than temperature gradients as driving agents for the velocity field. In liquids, $\varphi$ is a very large number, and thus incompressibility cannot be assumed even if $\nabla N / N$ may be small. Another consequence is that the sound velocity following from [(23),(104),(105)] increases with increasing $\varphi$, in agreement with the experimental fact that sound is significantly faster in liquids than in gases, and in distinctice disagreement with the sound velocity following from [(23),(60),(61)].

As a final remark: In the main body of this paper the emphasis has been on a rigorous treatment. Here, in Section 10, we have to be satisfied with less, owing to the immanent intricacy of the subject. Our present attempt to extend the applicability of the Navier--Stokes system to real fluids is accompanied by the hope that further work will be stimulated thereby. It would be unrealistic to expect that the Navier--Stokes equations in their standard form should be able to describe the dynamical behaviour of all fluids, ranging from ideal gases via real gases to liquids.

\vspace*{2cm}  

\centerline {\bf References}
\vspace*{3mm}

$^1$S.~Chapman and T.~Cowling, {\it The Mathematical Theory of Non--Uniform
Gases} (Cambridge University Press, 1953).

$^2$H.~Grad, ``Principles of the kinetic theory of gases,'' Handb.~Phys.~XII, 205 (1958). 

$^3$R.~Schunk, ``Mathematical structure of transport equations for multispecies flows,'' Rev.~Geophys.~Space Phys.~{\bf 15}, 429 (1977).

$^4$P.~Stubbe, ``The concept of a kinetic transport theory,'' Phys.~Fluids {\bf B 2}, 22 (1990).

$^5$P.~Constantin, ``On the Euler equation of incompressible fluids,'' Bull.~Amer.~Math.~Soc.~{\bf 44}, 603 (2007).

$^6$P.~Stubbe, ``A new collisional relaxation model for small deviations from equilibrium,'' J.~Plasma Phys.~{\bf 38}, 95 (1987).

$^7$E.~Gross and M.~Krook, ``Model for collision processes in gases,'' Phys.~Rev.~{\bf 102}, 593 (1956).

$^8$L.~Landau and E.~Lifshitz, {\it Course of Theoretical Physics, Vol.~6: Fluid Mechanics} (Pergamon Press, 1987).  

$^9$C.~Fefferman, ``Existence and smoothness of the Navier--Stokes equation'', http://www.claymath.org/
millennium/Navier-Stokes\_Equations/navierstokes.pdf

\end{document}